# Integrating optimal ridesharing matching into multimodal traffic model: Implications for policy and sustainable transport system

Yueqi Liu[a]   Ke Han[b,*]   Zhuoqian Yang[a]   Yanghong Yu[a]   Wen Ji[a]

[a]School of Transportation and Logistics, Southwest Jiaotong University, Chengdu, China
[b]School of Economics and Management, Southwest Jiaotong University, Chegndu, China

**ABSTRACT:** Integrating ridesharing matching explicitly into multimodal traffic models is crucial for accurately assessing the impacts of multimodal transport (MT) on urban economic and environmental aspects. This paper integrates an optimal ridesharing matching method into a path-based deterministic day-to-day traffic assignment framework, considers match cancellations, and captures the interactions between various modes on the road. The model incorporates five traffic modes (solo driving, ridesharing as a driver, ridesharing as a passenger, bus travel, and metro travel) and two groups of travelers based on their ownership status. Its steady state is determined through numerical experiments. The sensitivity analyses reveal that the MT system's performance varies with changes in ownership, bus fare, and ridesharing fare, demonstrating diverse impacts on mode split, travel cost, and emissions across different groups, road links, and regions. Our findings suggest that vehicle restrictions and pricing strategies have both benefits and drawbacks in managing MT system, emphasizing the need for careful consideration of trade-offs and social equity implications in policy-making and implementation. This study not only enhances the theoretical understanding of MT system but also provides valuable support for urban transportation policy-making aimed at achieving efficient, sustainable, and socially equitable transport systems.

## 1. Introduction

Ridesharing enables individuals with similar itineraries and time availability to share a vehicle (Furuhata et al., 2013). With the rise of smartphones and the increasing interest in mobility as a service (MaaS), multimodal transport (MT) with ridesharing has gained significant attention (Zhang et al., 2024). Existing studies show that incorporating ridesharing into multimodal transportation systems can reduce vehicle usage (Chan and Shaheen, 2012), alter trip frequency (Wang et al., 2019), and cause modal shift (Zhang et al., 2022, Hall et al., 2018), potentially alleviating congestion. However, it can also shift or exacerbate congestion (Beojone and Geroliminis, 2021, Nie, 2017). These effects also impact social welfare (Sun and Szeto, 2021), traffic-related air pollution (Brown, 2020), and greenhouse gas emissions (Tikoudis et al., 2021).

Research on MT model, specifically those integrating ridesharing into multimodal traffic assignment frameworks, is evolving. It helps facilitate the evaluation of MT system performance and derive policy implications such as congestion tolling, HOV lane design, and traffic restriction (Chen and Di, 2021, Di et al., 2017, Sun and Szeto, 2021). Despite advances since Davanzo's pioneering work, ridesharing matching is often simplified or omitted in models (Xu et al., 2015, Bahat and Bekhor, 2016, Ma et al., 2022, Pi et al., 2019, Ma and Zhang, 2017).For instance, many studies consider only the mode choice of ridesharing without accounting for the matching process (Pi et al., 2019), and some models address ridesharing matching as capacity constraints (Ma et al., 2020) or matching probabilities (Wei et al., 2020) without considering traveler behavioral patterns of driver and passenger (hereafter referred to as pax). Ridesharing matching is crucial for an accurate representation of ridesharing systems. Yao and Bekhor (2023) emphasize the necessity of explicitly integrating ridesharing matching into multimodal traffic models, yet few studies have incorporated this comprehensiveness. Current ridesharing matching systems are typically centralized, such as Uber and DIDI, and are often framed as optimization problems. These systems aim to maximize the number of matches (Najmi et al., 2017), minimize travel time (Naoum-Sawaya et al., 2015), travel miles (Nourinejad and Roorda, 2016), or emissions (Atahran et al., 2014). Recent studies have also employed machine learning techniques to enhance these algorithms(Qin et al., 2022).

Travelers continuously learn and adjust their behaviors based on daily experiences (Smith, 1984, Horowitz, 1984), which are modeled using day-to-day (DTD) traffic assignment models in multimodal networks (Cantarella and Fiori, 2022, Sun, 2023). These models, categorized into deterministic and stochastic processes or link-based and path-based approaches (Cheng et al., 2019, He et al., 2010, Guo et al., 2015, Huang and Lam, 2002, Friesz et al., 1994, Cantarella and Watling, 2016), help understand dynamic equilibrium (Yu et al., 2020) and the evolution of traffic flow through techniques like flow and perceptual updating (Huang and Lam, 2002, Yu et al., 2020, Cheng et al., 2019). Despite their value in providing critical decision support for traffic management and planning, the integration of DTD models with ridesharing matching in MT system remains limited. Existing studies either neglect route choices or treat public transit as independent of other modes(Wei et al., 2020, Yao and Bekhor, 2023), hindering policy analysis of network performance and behavioral evolution from an equity perspective.

To address these gaps, this paper proposes a day-to-day multimodal traffic model with optimal ridesharing matching, considering five traffic modes (solo driving, ridesharing as a driver, ridesharing as a pax, bus travel, and metro travel), two groups of travelers according to their ownership status, and travel paths. Within a path-based deterministic DTD framework, this model integrates optimal ridesharing matching, including ridesharing route generation and matching cancellation, and captures the interactions between various modes on the road, as well as modal shifts and route choices. By analyzing system performance under different policies in terms of modal split, time cost, monetary cost, and emissions, the model provides insights into efficiency, environmental impact, and equity. This approach not only enhances the integration of ridesharing with multimodal traffic models but also offers valuable support for urban transportation policy-making. The contributions of this study are outlined below:

Novel integration of ridesharing matching in multimodal traffic assignment models: This paper presents a path-based deterministic DTD traffic assignment model that incorporates ridesharing matching optimization, addressing interactions between different transportation modes and groups of travelers' behaviors. It offers a comprehensive approach to multimodal traffic modeling.

Detailed policy impact analysis: The proposed model facilitates an in-depth analysis of system performance changes under various policies. By employing model sensitivity analysis, the study provides valuable insights into the efficiency, environmental impact, and equity of policies such as vehicle restrictions and pricing strategies.

Enhanced support for urban transportation policy-making: By combining ridesharing matching with a deterministic DTD framework, this paper offers a new methodology for integrating ridesharing into multimodal traffic assignment models. This approach enhances both the theoretical understanding and practical application of multimodal transport systems, supporting effective policy development across various transportation modes.

The rest of the paper is organized as follows. Section 2 describes the overall framework and the multimodal traffic model we proposed. Section 3 conducts the numerical experiments for empirical convergence. Section 4 performs sensitivity analysis. Section 5 discusses policy implications. Finally, Section 6 concludes this paper and outlines further work.

## 2. Method

### 2.1. Overall framework

The multimodal network in this study encompasses solo driving, ridesharing, and public transportation, involving vehicles like small cars, buses, and metro. Travelers are classified into two groups based on their ownership: vehicle owners (referred to as 'owners') and non-owners, denoted as $g \in G = \{0,1\}$. Consequently, the network comprises five distinct travel modes: solo driving, ridesharing as a driver, ridesharing as a pax, bus travel, and metro travel, denoted as $m \in M = \{1,2,3,4,5\}$. As indicated in Table 1, a traveler owning a car can opt to be either a driver or a pax, selecting from all travel modes, i.e., $m \in \{1,2,3,4,5\}$ for $g = 1$. Additionally, non-owners are restricted to being pax and can only utilize ridesharing, bus, or metro services, i.e., $m \in \{3,4,5\}$ for $g = 0$.

Table 1 Travelers' groups and travel modals

|  | Driver | | Pax | | |
| --- | --- | --- | --- | --- | --- |
|  | Solo driving ($m = 1$) | Ridesharing driving ($m = 2$) | Ridesharing taking ($m = 3$) | Bus taking ($m = 4$) | Metro taking ($m = 5$) |
| Travelers who own vehicles ($g = 1$) | $(g, m) = (1, 1)$ | $(g, m) = (1, 2)$ | $(g, m) = (1, 3)$ | $(g, m) = (1, 4)$ | $(g, m) = (1, 5)$ |
| Travelers who do not own vehicles ($g = 0$) | -- | -- | $(g, m) = (0, 3)$ | $(g, m) = (0, 4)$ | $(g, m) = (0, 5)$ |

The multimodal traffic modeling approach and policy analysis process of this study are illustrated in Figure 1. Each day, travelers would choose from one of the five aforementioned traffic modes. Among these, those who opt for ridesharing announce their requests in advance and will be matched by a centralized matching process, details shown in section 2.2.1. Once travel modes are determined, the resulting flows of private cars, ridesharing vehicles, and buses interact on the road network, creating a mixed traffic flow, while the subway operates on an independent right-of-way. The within-day actual travel costs form the basis for perceived costs, and the modal and route choices are updated day to day using by the perceived cost and logit models. We then introduce disturbances to the system and conduct sensitivity analysis to observe changes in system performance, thereby deriving policy implications.

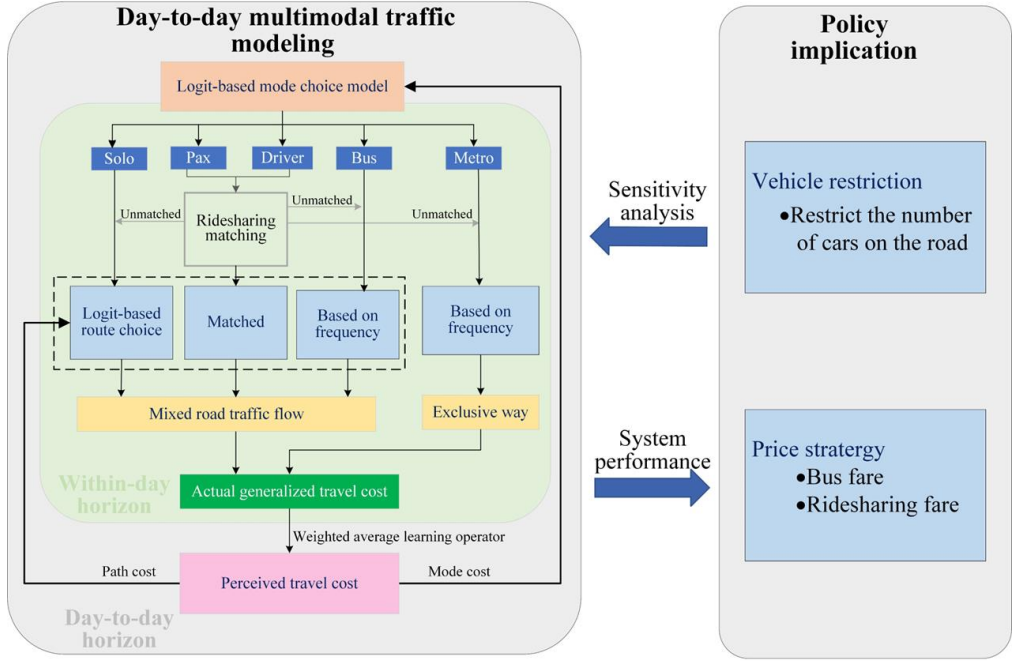

**Figure 1 Overall framework of the day-to-day multimodal traffic modeling and policy analysis process**

## 2.2. Within-day matching and multimodal cost

### 2.2.1. Ridesharing matching mechanism

#### 2.2.1.1. Centralized optimal matching

This paper employs a centralized static ridesharing matching approach in which the demands of drivers and passengers are predetermined or pre-announced. The pre-announced information includes the role (driver or pax) and the trip's origin and destination (OD). To be designated as a driver, an individual must own a car, whereas pax can either own a car or not. Let $KK$ denote the set of announced drivers and $PP$ represent the set of announced pax, where $KK = \{kk \in KK | g_{kk} = 1\}$ and $PP = \{pp \in PP | g_{pp} \in \{0,1\}\}$. The total ridesharing demand is $KK \cup PP$. This paper focuses on scenarios where only one driver is taking one pax per match, and the driver proceeds to their destination after dropping off the pax.

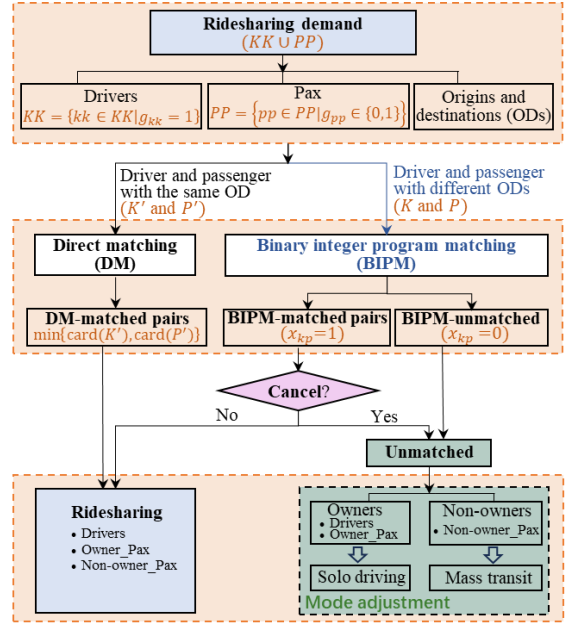

**Figure 2 Ridesharing matching process**

The process of ridesharing matching primarily comprises the announcement of ridesharing demand, centralized matching, cancellation, and mode adjustment. To expedite matching computation and enhance the overall matching success rate, two types of matching are established based on whether the driver and pax have the same origin and destination, as shown in Figure 2: direct matching (DM), followed by binary integer program matching (BIPM). In the DM process, drivers and pax have the same origin and destination (i.e., $O_{kk} = O_{pp}$, and $D_{kk} = D_{pp}$). The sets of such drivers and pax are denoted as $K'$ and $P'$, respectively. Obviously, the number of DM-matched pairs is the minimum of the sizes of sets $K'$ and $P'$, expressed as $\min\{\text{card}(K'), \text{card}(P')\}$. The remaining drivers and pax who do not have the same origin and destination are denoted as $K = \{k \in K | g_k = 1\}$ and $P = \{p \in P | g_p \in \{0,1\}\}$, respectively, thus $K \cup K' = KK$ and $P \cup P' = PP$. Based on Agatz et al. (2011) and Alisoltani et al. (2021), the BIPM is developed for the remaining travelers: The objective function aims to maximize the total in-vehicle time savings (TTS), with in-vehicle time derived in Section 2.2.2.1. The decision variable $x_{kp}$ equals 1 if driver $k$ and pax $p$ are matched, namely BIPM-matched pairs, and 0 otherwise, namely BIPM-unmatched travelers. Matched trips consistently utilize the shortest route between origin and destination nodes, with routes dynamically changing daily.

$$\max TTS = \sum_{k \in K} \sum_{p \in P} x_{kp} \cdot S_{kp} \tag{2.1}$$

$$S_{kp} = T_{O_k D_k} + T_{O_p D_p} - \left(T_{O_k O_p} + T_{O_p D_p} + T_{D_p D_k}\right), \quad \forall k \in K, p \in P \tag{2.2}$$

$$\sum_{p \in P} x_{kp} \leq 1, \quad \forall k \in K \tag{2.3}$$

$$\sum_{k \in K} x_{kp} \leq 1, \quad \forall p \in P \tag{2.4}$$

$$x_{kp} = \{1,0\}, \quad \forall k \in K, p \in P \tag{2.5}$$

in which, Eq. (2.2) denotes the disparity between the total original travel time of the participants and the revised travel time after matching. Constraints (2.3) and (2.4) dictate that each driver (pax) can be matched with, at most, one pax (driver).

### 2.2.1.2. Cancellation and mode adjustment

By solving the BIPM, we can obtain BIPM-matched pairs and BIPM-unmatched travelers. Therefore, announcers in DM-matched pairs and those accepting BIPM-matched will travel by ridesharing, as shown in Figure 2. However, considering that excessive pick-up and drop-off distances potentially prompt cancellation of the centralized matched pairs (Furuhata et al., 2013), this paper posits that ridesharing drivers will cancel the BIPM-matched trip when the empty mileage $d_{cancel}$ surpasses a specific threshold, denoted as $d_{cancel} > threshold$, forcing the pax paired with these drivers to also cancel the match. Those who cancel the BIPM-matched, along with the BIPM-unmatched travelers, are all unmatched. Ultimately, all the unmatched announcers will adjust to alternative modes. There are many factors that ridesharing announcers consider when adjusting their travel modes (Zhang et al., 2024), this paper focuses on the inertia towards vehicle use. It assumes that car owners are more likely to continue using private cars when adjusting their modes. Consequently, unmatched drivers and car-owning pax will switch to driving alone, while unmatched pax without cars will choose between the metro and buses based on cost.

### 2.2.2. Multimodal generalized cost

#### 2.2.2.1. In-vehicle time

The Bureau of Public Roads (BPR) link cost function is employed to calculate the in-vehicle time, capturing the road congestion. Cars (used by solo driving and ridesharing) and buses are considered to share the same road, resulting the heterogeneous traffic flow $f_a$. Therefore, the in-vehicle time $t_a$ spent traveling on link $a$ is given by:

$$t_a = t_a^0(1 + \alpha_1(f_a/q_a)^{\alpha_2}) \qquad (2.6)$$

where $t_a^0$ is the free-flow time cost of link $a$. $\alpha_1, \alpha_2$ are the calibration parameters. $q_a$ is the road capacity of link $a$. $f_a$ is the heterogeneous traffic flow of link $a$ (unit: pcu/h), excluding metro, which operates on its own right-of-way. The traffic flow requires the conversion of the departure trip flow:

$$f_a = \sum_{w \in W} \sum_{g \in G} \sum_{r(s) \in R(s)} \delta_{a,r(s)}^w h_{r(s),m=1}^{w,g} + \sum_{w \in W} \sum_{g \in G} \sum_{r(rd) \in R(rd)} \delta_{a,r(rd)}^w h_{r(rd),m=2}^{w,g} + \sum_{lb \in LB} \delta_{a,lb} \rho_{lb} \gamma \qquad (2.7)$$

where the first two items on the left represent the number of cars utilized by solo driving and ridesharing, It assumes that the number of cars is equal to the number of the drivers, that is, $h_{r(s),m=1}^{w,g} = hp_{r(s),m=1}^{w,g}$, and $h_{r(rd),m=2}^{w,g} = hp_{r(rd),m=2}^{w,g}$. The third item represents the number of buses converted to car-equivalent, based on the frequency $\rho_{lb}$ of bus line $lb$, and $\gamma$ is the conversion factor for buses compared to car-equivalents. Meanwhile,

$$\delta_{a,r(s)}^w = \begin{cases} 1, \text{ if route } r(s) \text{ for OD pair } w \text{ uses link } a \\ 0, \text{ otherwise} \end{cases}$$

$$\delta_{a,r(rd)}^w = \begin{cases} 1, \text{ if route } r(rd) \text{ for OD pair } w \text{ uses link } a \\ 0, \text{ otherwise} \end{cases}$$

$$\delta_{a,lb} = \begin{cases} 1, \text{ if bus line } lb \text{ uses link } a \\ 0, \text{ otherwise} \end{cases}$$

#### 2.2.2.2. Generalized cost

The generalized travel costs for different modes on their respective routes are defined.
1. For solo driving:

The generalized cost on route $r(s)$ of solo driving includes in-vehicle time cost, fixed cost, and fuel cost, represented by:

$$C_{r(s),m=1}^{w,g} = t_{r(s),m=1}^{w}\eta_{VOT} + c_{fixed} + c_{fuel}d_{r(s),m=1}^{w} \tag{2.8}$$

$$t_{r(s),m=1}^{w} = \sum_{a \in A} \delta_{a,r(s)}^{w} t_a(f_a) \tag{2.9}$$

$$d_{r(s),m=1}^{w} = \sum_{a \in A} \delta_{a,r(s)}^{w} d_a \tag{2.10}$$

where $c_{fuel}$ is the fuel cost per unit distance. $d_a$ is the distance of road link $a$. $\eta_{VOT}$ is the coefficient of the value of time.

2. For ridesharing driving:

The generalized cost of ridesharing driving on route $r(rd)$ includes in-vehicle time cost, fixed cost, fuel cost, matching cost, and the pax's ridesharing fare which is proportional to the distance traveled by the matched pax. And it assumes that the matching cost is a certain rate $\beta$ of the pax's fare. The cost for ridesharing driving is described as:

$$C_{r(rd),m=2}^{w,g} = t_{r(rd),m=2}^{w}\eta_{VOT} + c_{fixed} + c_{fuel}d_{r(rd),m=2}^{w} + (1-\beta)c_{FARE} \tag{2.11}$$

$$t_{r(rd),m=2}^{w} = \sum_{a \in A} \delta_{a,r(rd)}^{w} t_a(f_a) \tag{2.12}$$

$$d_{r(rd),m=2}^{w} = \sum_{a \in A} \delta_{a,r(rd)}^{w} d_a \tag{2.13}$$

$$c_{FARE} = c_{r-fare} \sum_{a \in A} \delta_{a,r(rp)}^{w} d_a \tag{2.14}$$

$$\delta_{a,r(rp)}^{w} = \begin{cases} 1, \text{ if route } r(rp) \text{ for OD pair } w \text{ uses link } a \\ 0, \text{ otherwise} \end{cases}$$

where $c_{r-fare}$ is the ride fare per unit distance.

3. For ridesharing taking:

Considering in-vehicle time, ridesharing waiting time $t_{rp-waite}$, and ridesharing fare, the generalized cost of taking ridesharing on route $r(rp)$ is given by:

$$C_{r(rp),m=3}^{w,g} = (t_{r(rp),m=3}^{w} + t_{rp-waite})\eta_{VOT} + c_{FARE} \tag{2.15}$$

$$t_{r(rp),m=3}^{w} = \sum_{a \in A} \delta_{a,r(rp)}^{w} t_a(f_a) \tag{2.16}$$

4. For bus taking:

Traveling by bus incurs time costs, including in-vehicle time, waiting time, and walking time to and from the stop. The in-vehicle time for buses is assumed to be $\sigma$ times that of cars when sharing the same road (Kawakami and Shi, 1994). The average waiting time for public transport is typically considered to be half of the frequency (Wei et al., 2020). Additionally, it assumes that pax must fare pay a flat bus fare here. Hence, the generalized cost of the bus on route $r(lb)$ is shown by:

$$C_{r(lb),m=4}^{w,g} = \left(t_{r(lb),m=4}^{w} + \frac{1}{2\rho_{lb}} + t_{b-walk}\right)\eta_{VOT} + c_{b-fare} \tag{2.17}$$

$$t_{r(lb),m=4}^{w} = \sigma \sum_{a \in A} \delta_{a,r(lb)} t_a(f_a) \tag{2.18}$$

$$\delta_{a,r(lb)} = \begin{cases} 1, \text{if route } r(lb) \text{ for OD pair } w \text{ uses link } a \\ 0, \text{ otherwise} \end{cases}$$

5. For metro taking:

The time cost of taking metro comprises the same components as taking the bus. Consequently, the generalized cost of the metro on route $r(lm)$ is:

$$C_{r(lm),m=5}^{w,g} = \left( t_{r(lm),m=5}^{w} + \frac{1}{2\rho_{lm}} + t_{me-walk} \right) \eta_{VOT} + c_{me-fare} \tag{2.19}$$

Finally, the average modal travel cost for mode $m$ is obtained by weighting its route flow with the generalized route cost:

$$C_m^{w,g} = \sum_r hp_{r,m}^{w,g} \cdot C_{r,m}^{w,g} \bigg/ \sum_r hp_{r,m}^{w,g} \tag{2.20}$$

## 2.3. Day-to-day cost and travel choice updating mechanism

### 2.3.1. Perceived cost

DTD Updating techniques are mainly categorized into flow updating and perceptual updating. Flow updating involves directly setting a traffic flow update rule, such as adjusting based on a certain percentage (e.g., Huang and Lam, 2002). Perceptual updating views the evolution of network traffic flow as a consequence of the traveler's perceived cost and subsequent learning, where the perceived cost can be expressed as a linear combination of the experienced and actual costs (e.g., Yu et al., 2020, Cheng et al., 2019). In this paper, the weighted average learning operator (Yu et al., 2020, Cascetta, 1989) is used to express the perceived mode cost and perceived path cost. The perceived mode cost on day $\tau$ is formulated as:

$$\overline{C}_m^{w,g}(\tau) = \frac{1}{\sum_{i=1}^{N} \lambda^{i-1}} \left( C_m^{w,g}(\tau-1) + \lambda C_m^{w,g}(\tau-2) + \cdots + \lambda^{N-1} C_m^{w,g}(\tau-N) \right) \tag{2.21}$$

where $\lambda \in (0,1)$, represents the weight of the past days' experienced costs. $N$ is the memory days which is the number of past days that influence the present day's decision. Similarly, the perceived path cost of mode $m$ on day $\tau$ could be:

$$\overline{C}_{r,m}^{w,g}(\tau) = \frac{1}{\sum_{i=1}^{N} \lambda^{i-1}} \left( C_{r,m}^{w,g}(\tau-1) + \lambda C_{r,m}^{w,g}(\tau-2) + \cdots + \lambda^{N-1} C_{r,m}^{w,g}(\tau-N) \right) \tag{2.22}$$

### 2.3.2. Logit model

Following the random utility theory, we define the expected travel cost as the sum of the perceived cost and a random term:

$$\hat{C}_m^{w,g}(\tau) = \overline{C}_m^{w,g}(\tau) + \varepsilon_m^{w,g} \tag{2.23}$$

$$\hat{C}_{r,m}^{w,g}(\tau) = \overline{C}_{r,m}^{w,g}(\tau) + \varepsilon_{r,m}^{w,g} \tag{2.24}$$

where $\hat{C}_m^{w,g}(\tau)$ and $\hat{C}_{r,m}^{w,g}(\tau)$ represent the expected mode cost and expected path cost, respectively. When random term $-\varepsilon_m^{w,g}, -\varepsilon_{r,m}^{w,g}$ are independent and identically Gumbel distributed, the probability of choosing mode $m$ on day $\tau$ can be determined based on the logit model:

$$P_m^{w,g}(\tau) = \frac{exp(-\theta_1 \overline{C}_m^{w,g}(\tau))}{\sum_{(m')} exp(-\theta_1 \overline{C}_{m'}^{w,g}(\tau))} \quad \forall (m') \neq (m), \theta_1 > 0 \tag{2.25}$$

In addition, we assume that travelers make decisions about traveling mode and route sequentially. Therefore, the probability of choosing mode $m$ with route $r$ is expressed as the product of the mode choice probability $P_m^{w,g}(\tau)$ and conditional probability $P_{(r|m)}^{w,g}(\tau)$. Since the routes for ridesharing are determined by the matching process, and the routes for buses and metro are fixed, only the probability of choosing paths of solo driving ($m=1$) is given by the logit model:

$$P_{(r|m=1)}^{w,g}(\tau) = \frac{exp(-\theta_2 \overline{C}_{r,m=1}^{w,g}(\tau))}{\sum_{(r')} exp(-\theta_2 \overline{C}_{r,m=1}^{w,g}(\tau))} \quad \forall r' \neq r, \theta_2 > 0 \tag{2.26}$$

Hence, the departure trip flow of different modes could be expressed as:
$$hp_m^{w,g}(\tau) = Hp^{w,g} \cdot P_m^{w,g}(\tau) \tag{2.27}$$

where $hp_m^{w,g}(\tau)$ is the modal departure trip flow, stranding for the mode choice. $Hp^{w,g}$ is the total departures all of modes. Note that $hp_m^{w,g}(\tau)$ will be adjusted and updated on day $\tau$ after the ridesharing match confirmation. Furthermore, the modal path flow of solo driving ($m=1$) with route $r$ could be formulated:

$$hp_{r,m=1}^{w,g}(\tau) = Hp^{w,g} P_{m=1}^{w,g}(\tau) P_{(r|m=1)}^{w,g}(\tau), \quad r = r(s) \tag{2.28}$$

Meanwhile, as for the modal path flow of ridesharing, i.e., $hp_{r,m=2}^{w,g}(\tau)$ and $hp_{r,m=3}^{w,g}(\tau)$, they can be obtained by loading the modal departure trip flow $hp_{m=2}^{w,g}(\tau)$, $hp_{m=3}^{w,g}(\tau)$ to the matched path output by the matching algorithm. And for bus and metro, i.e., $hp_{r,m=4}^{w,g}(\tau)$ and $hp_{r,m=5}^{w,g}(\tau)$, they are equal to the bus and metro trips on the already fixed public transportation routes between OD pair $w$.

*2.3.3. Steady state*

The DTD modeling reaches a stable state where the daily traffic distribution does not experience significant fluctuations, achieving a form of dynamic equilibrium (Yu et al., 2020).The constancy of path flows from day to day implies the traffic equilibrium in the path-based DTD modeling(Yang and Zhang, 2009), which is expected to satisfy the following fixed-point conditions (Liu and Geroliminis, 2017, Wei et al., 2020, Cantarella and Fiori, 2022, Ye et al., 2021, Liu et al., 2024a):

$$\overline{\boldsymbol{C}}^* = \boldsymbol{C}(\boldsymbol{hp}^*), \boldsymbol{hp}^* = \Lambda(\overline{\boldsymbol{C}}^*) \tag{2.29}$$

where $\Lambda(\bullet)$ is network loading function. Perceived cost is equal to experienced (actual) cost and remains constant at the fixed point, showing that $\overline{C}_m^{w,g}(\tau+1) = \overline{C}_m^{w,g}(\tau) = C_m^{w,g}(\tau)$, and $\overline{C}_{r,m}^{w,g}(\tau+1) = \overline{C}_{r,m}^{w,g}(\tau) = C_{r,m}^{w,g}(\tau)$, here $hp_m^{w,g}(\tau+1) = hp_m^{w,g}(\tau) = hp_m^{w,g,*}$, and $hp_{r,m}^{w,g}(\tau+1) = hp_{r,m}^{w,g}(\tau) = hp_{r,m}^{w,g,*}$.

*2.4. Emission calculation*

In addition to analyzing the modal split and traffic conditions of MT system, this paper also estimates the emissions of air pollutants and greenhouse gases (GHGs), given their increasing attention due to health and environmental concerns (Bai et al., 2022, Liu et al., 2024b, Andress et al., 2011). Specifically, this paper considers air pollutant emissions including $NO_X$, $PM_{2.5}$, CO, and GHGs emissions in terms of carbon dioxide equivalent (CO2e). Based on the emission factor, the vehicle emissions is calculated as Zou et al. (2023) and Zhao et al. (2016):

$$E_a^{z_1} = \sum_m EF_m^{z_1} \cdot f_{m,a} \cdot d_a, \quad m = \{1,2,4\} \qquad (2.30)$$

$$E_a^{z_2} = \sum_m EF_m^{z_2} \cdot fuel \cdot f_{m,a} \cdot d_a, \quad m = \{1,2,4\} \qquad (2.31)$$

where $E_a^{z_1}, E_a^{z_2}$ represent the emission of air pollutants $z_1 = \{NO_X, PM_{2.5}, CO\}$ and $z_2 = \{CO2e\}$ of the link $a$, respectively (unit: g/h). $f_{m,a}$ is the traffic flow of mode $m$ of the link $a$ (unit: veh/h). $d_a$ is the distance of link $a$ ((unit: km). $EF_m^{z_1}, EF_m^{z_2}$ denote the emission factors of $z_1$ and $z_2$ for the vehicle of mode *m*, respectively. Here the unit of $EF_m^{z_1}$ is g/km/veh and the unit of $EF_m^{z_2}$ is g/L/veh. Lastly, fuel consumption is denoted as $fuel$ (unit: L/km).

## 3. Numerical experiments

*3.1. Experimental setup*

The numerical experiment is conducted on the Sioux-Falls network, comprising 24 nodes and 76 links. We implement two bus lines and two metro lines to facilitate north-south traffic flow within the network, as depicted in Figure 3. Buses share the road with cars, while the metro operates on a dedicated right-of-way. The OD set contains 20 OD pairs, 13 of which have access to mass transit services. The OD demand and parameter values are provided in Figure 3 and Table 2, respectively. Notably, monetary costs and vehicle operation parameters are adapted to the Chinese urban context. For example, the fuel cost per unit distance for cars is approximately 0.5 yuan/km (Guo et al., 2021). Cars are assumed to be powered by gasoline, while mass transit vehicles are considered fully electrified, with no on-road emissions. The priori route sets for the cars derive from the Frank-Wolfe algorithm (Yu et al., 2020) and Dijkstra's algorithm is employed to compute the shortest time-cost path. The Gorubi solver is used to solve the integer programming model for matching.

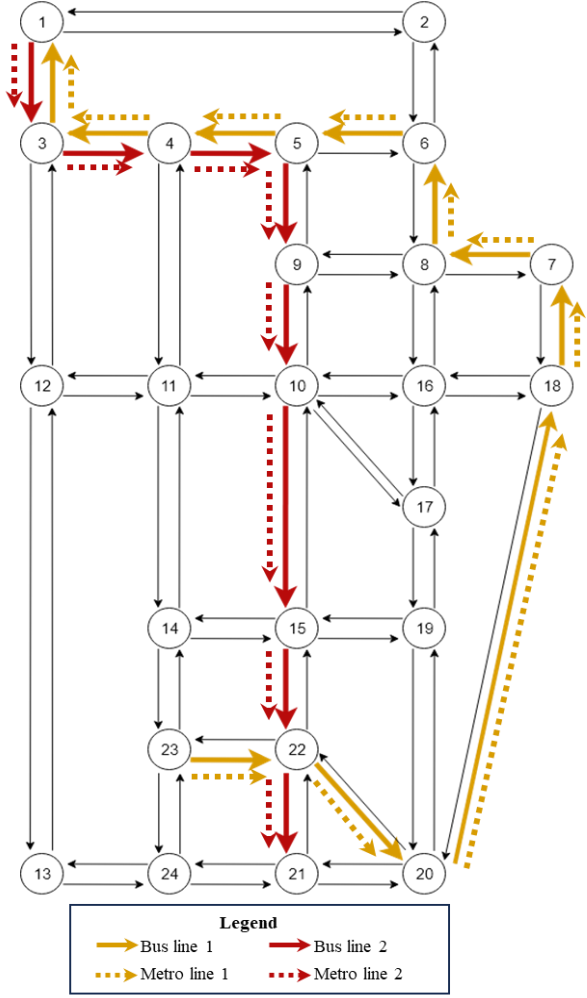

Demand setting of the Sioux-Falls network example

| O | D | Own-vehicle | | | | | |
|---|---|---|---|---|---|---|---|
| | | Demand (trip /h) | % Solo | % Pax | % Driver | % Bus | % Metro |
| 20 | 13 | 1470 | 90 | 0 | 10 | 0 | 0 |
| 22 | 17 | 2100 | 70 | 30 | 0 | 0 | 0 |
| 4 | 1 | 1680 | 30 | 30 | 20 | 10 | 10 |
| 22 | 19 | 210 | 70 | 0 | 30 | 0 | 0 |
| 16 | 2 | 630 | 90 | 10 | 0 | 0 | 0 |
| 3 | 9 | 1890 | 50 | 10 | 20 | 10 | 10 |
| 7 | 16 | 2100 | 50 | 40 | 10 | 0 | 0 |
| 13 | 24 | 840 | 80 | 20 | 0 | 0 | 0 |
| 23 | 15 | 420 | 50 | 20 | 30 | 0 | 0 |
| 23 | 18 | 1470 | 50 | 30 | 0 | 10 | 10 |
| 4 | 18 | 630 | 70 | 20 | 10 | 0 | 0 |
| 23 | 20 | 210 | 40 | 20 | 20 | 10 | 10 |
| 23 | 11 | 2100 | 40 | 30 | 30 | 0 | 0 |
| 12 | 9 | 420 | 80 | 0 | 20 | 0 | 0 |
| 19 | 17 | 1050 | 70 | 30 | 0 | 0 | 0 |
| 4 | 9 | 2100 | 40 | 0 | 40 | 10 | 10 |
| 10 | 21 | 210 | 50 | 20 | 10 | 10 | 10 |
| 22 | 7 | 420 | 60 | 0 | 20 | 10 | 10 |
| 19 | 10 | 2100 | 60 | 10 | 30 | 0 | 0 |
| 23 | 6 | 420 | 50 | 10 | 20 | 10 | 10 |

| O | D | Non-own vehicle | | | | Mass transit | |
|---|---|---|---|---|---|---|---|
| | | Demand (trip /h) | % Pax | % Bus | % Driver | Bus line | Metro line |
| 20 | 13 | 0 | 0 | 0 | 0 | 0 | 0 |
| 22 | 17 | 0 | 0 | 0 | 0 | 0 | 0 |
| 4 | 1 | 720 | 20 | 40 | 40 | 2 | 2 |
| 22 | 19 | 0 | 0 | 0 | 0 | 0 | 0 |
| 16 | 2 | 0 | 0 | 0 | 0 | 0 | 0 |
| 3 | 9 | 810 | 20 | 40 | 40 | 1 | 1 |
| 7 | 16 | 0 | 0 | 0 | 0 | 0 | 0 |
| 13 | 24 | 0 | 0 | 0 | 0 | 0 | 0 |
| 23 | 15 | 0 | 0 | 0 | 0 | 0 | 0 |
| 23 | 18 | 630 | 20 | 40 | 40 | 2 | 2 |
| 4 | 18 | 0 | 0 | 0 | 0 | 0 | 0 |
| 23 | 20 | 90 | 20 | 40 | 40 | 2 | 2 |
| 23 | 11 | 0 | 0 | 0 | 0 | 0 | 0 |
| 12 | 9 | 0 | 0 | 0 | 0 | 0 | 0 |
| 19 | 17 | 0 | 0 | 0 | 0 | 0 | 0 |
| 4 | 9 | 900 | 20 | 40 | 40 | 1 | 1 |
| 10 | 21 | 90 | 20 | 40 | 40 | 1 | 1 |
| 22 | 7 | 180 | 20 | 40 | 40 | 2 | 2 |
| 19 | 10 | 0 | 0 | 0 | 0 | 0 | 0 |
| 23 | 6 | 180 | 20 | 40 | 40 | 2 | 2 |

**Figure 3 Sioux-Falls network with public transport and the demand setting**

Table 2 Parameter values in the example

| Index | Notations | Values |
|---|---|---|
| Value of time (yuan/min) | $\eta_{VOT}$ | 0.3 |
| Coefficient in BPR function | $\alpha_1, \alpha_2$ | 0.15, 4 |
| Gasoline cost per unit distance of cars (yuan/km) | $c_{fuel}$ | 0.5 |
| Unit fuel consumption of gasoline (L/km) | $fuel$ | 0.05 |
| Fixed cost of car (yuan) | $c_{fixed}$ | 1 |
| The charge of ridesharing (yuan/km) | $c_{r-fare}$ | 2 |
| Payments for public transit (yuan) | $c_{b-fare}, c_{me-fare}$ | 5, 5 |
| Frequency of public transit (veh/h) | $\rho_{lb}, \rho_{lm}$ | 20, 20 |
| Conversions factor between car and bus | $\gamma, \sigma$ | 4.5, 1.5 |
| Walking time for public transit [a] (min) | $t_{b-walk}, t_{me-walk}$ | 7, 11 |
| Coefficient in Logit model | $\theta_1, \theta_2$ | 0.004, 0.004 |
| Emission factor [b] | $EF_{m=1}^{z_1}, EF_{m=1}^{z_2}$ | CO=0.46, NOx=0.017, PM2.5=0.003, CO2e=2360 |
| Matching cost rate | $\beta$ | 0 [c] |
| Days of memory | $N$ | 30 |
| Weights of past days' experienced costs | $\lambda$ | 0.7 |
| Threshold for ridesharing driver to cancel the BIPM-matched (km) | | 10 |

a: Using the service radius of 500m for buses and 800m for subways, and assuming a walking speed of 1.2m/s, the average walking time within the service area is calculated.
b: Emission factor of CO, $NO_X$, $PM_{2.5}$ is based on the National V standard emission of small cars (MEPC, 2015). And emission factor of $CO_2e$ is based on Zhao et al. (2016)
c: This study analyzes the impacts of ownership, bus fares, and ridesharing fares. For simplicity, the matching cost rate is set to 0 in the numerical study, which does not affect the analysis results.

### 3.2. Empirical convergence

We utilize two metrics to monitor the oscillatory behavior of the system and assess the convergence of experimental results. These metrics are: (1) the daily disparity between perceived and experienced (actual) costs of various modes, referred to as $gap_1$ (Liu and Geroliminis, 2017, Yu et al., 2020), and (2) the variation in flows between two consecutive days, denoted as $gap_2$. They are expressed as:

$$gap_1 = \frac{\left|\overline{C}_m^{w,g}(\tau) - C_m^{w,g}(\tau)\right|}{\left|\overline{C}_m^{w,g}(\tau)\right|} \quad (3.1)$$

$$gap_2 = \frac{\left|h_m^{w,g}(\tau+1) - h_m^{w,g}(\tau)\right|}{h_m^{w,g}(\tau)} \quad (3.2)$$

where $gap_2$ reflects the day-to-day changes in the mode split, indicating variations in the daily experienced (actual) costs. When both of the gaps approach 0, the fixed point is considered satisfied. To assess convergence under varied initial conditions, we simulated different scenarios where $N$ =3, 6, or 30 with $\theta_1 = \theta_2$ equals 0.004, and $\theta_1 = \theta_2$=0.01, 0.004, or 0.001 with $N$ equals 30. The final errors are all less than $10^{-4}$, suggesting that the developed model achieved a stable state in the numerical experiments, as depicted in Figure 4. Moreover, this demonstrates that increasing $N$ can mitigate the daily fluctuations, particularly when $\theta$ is small.

The convergence of the proposed multimodal traffic model as well as the uniqueness of the solution can be affected by the cost function and the network condition, and is especially prone to fail in multi-modal traffic in large-scale networks. Meanwhile, as mentioned in Liu and Geroliminis (2017) and Yu et al. (2020), it's crucial to acknowledge that day-to-day convergence may perform poorly when travelers are highly sensitive to cost changes (e.g. when $N$ is small or $\theta$ is large). We defer these intriguing yet challenging issues for future investigation. While this paper does not delve deeply into these problems, we present daily evolution and its eventual convergence through numerical experiments, further conducting sensitivity analysis. This expands the modeling of multimodal transportation networks with ridesharing, offering a deeper understanding of the MT system and insights for urban management.

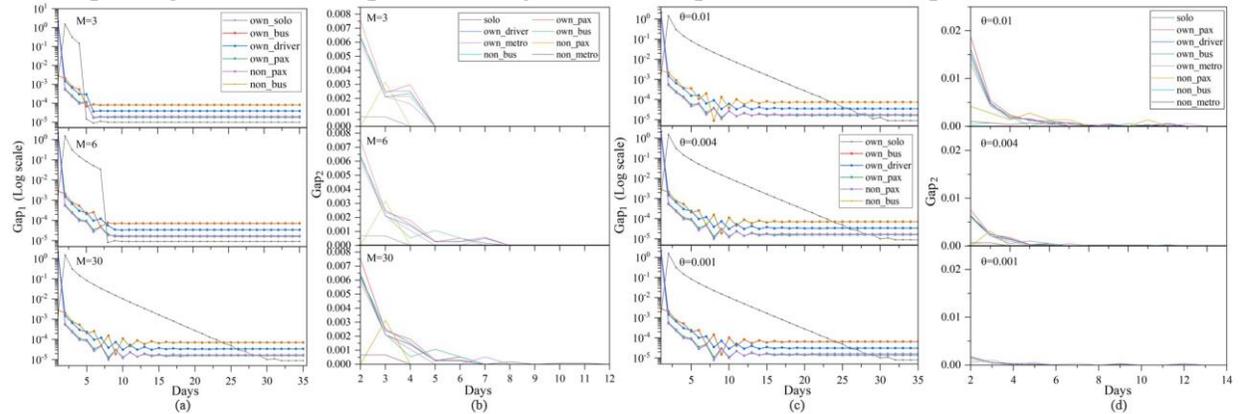

**Figure 4 Convergence test under different initial conditions with $N$= 3, 6, 30 ($\theta_1 = \theta_2$=0.004), and $\theta_1 = \theta_2$=0.01, 0.004, 0.001 ($N$=30)**

## 4. Sensitivity analysis

We analyze the system performance and conduct sensitivity analysis with $N=30$, $\theta_1 = \theta_2 = 0.004$, assessing mode split and PCU, travel cost, and emissions. Disturbances in ownership, bus fare, and ridesharing fare are also examined and compared to the base condition without disturbances, offering insights into policy implications. Analyses are performed separately for all ODs and mass-transit ODs to assess system interference from both global and local perspectives.

### 4.1. Vehicle restriction[1]

Car ownership on mass-transit ODs is banned at 30%, 50%, and 70%, denoted by ban=0.3, 0.5, and 0.7, respectively. And ban=0 indicates no restrictions. Consequently, restricted car owners will be considered non-owners and can only take public transit or ridesharing as pax. Below are the comparisons of the stable state for different ownership scenarios:

Table 3 illustrates the modal split and PCU changes as ownership decreases. Both all-ODs and mass-transit ODs experience reduced solo driving and ridesharing, along with increased mass transit use, resulting in lower total PCUs on the road. Additionally, Figure 5 details travel behavior across ownership groups. For car owners, ownership restrictions lead to fewer trips via solo driving, ridesharing, or mass transit. For example, with a 70% ownership ban, solo driving decreases by 6% for all ODs (Figure 5 (b)). Similarly, mass transit use drops from 14% to 4%, and ridesharing decreases from 28% to 21%. Non-owners, however, show increased public transit demand, with ridesharing trends depending on ownership reduction. With a 30% ownership cut (ban=0.3), the number of non-owners increases, and their ridesharing usage rises from 2% to 3%, as illustrated in Figure 5 (b). Conversely, at 70% ((ban=0.7), despite a larger increase in non-owners, ridesharing decreases by 1%, constrained by driver availability. Figure 5 (a) shows the trends in travel behavior from mass-transit ODs mirror those from all ODs, but are more pronounced. For example, if ban=0.7, the proportion of non-owner trips using public transit increases by 52%, while ridesharing decreases by 2%, compared to a 22% increase and 1% decrease for all ODs.

Table 3  Modal split under different ownership bans

| Ban | All ODs | | | | Mass-transit ODs | | | |
| --- | --- | --- | --- | --- | --- | --- | --- | --- |
| | Solo | Mass transit (bus & metro) | Ridesharing | PCU | solo | Mass transit (bus & metro) | Ridesharing | PCU |
| 0 | 11359 | 6714 | 7998 | 15448 | 2271 | 6714 | 3014 | 3868 |
| 0.3 | 10906 | 8070 | 7094 | 14543 | 1818 | 8070 | 2110 | 2963 |
| 0.5 | 10475 | 9105 | 6492 | 13811 | 1387 | 9105 | 1508 | 2231 |
| 0.7 | 9968 | 10217 | 5890 | 13003 | 880 | 10217 | 906 | 1423 |

---

[1] In this analysis, while we nominally adjust "ownership", we are in fact controlling the number of cars on the road, corresponding to real-world vehicle restriction policies.

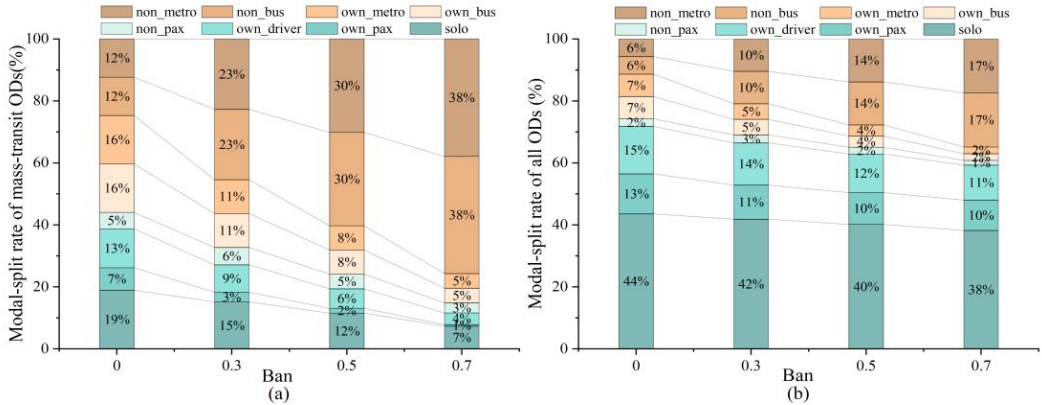

**Figure 5    Modal-split rate of different groups under bans**

For time costs and monetary costs, as depicted in Figure 6, the average in-vehicle time decreases with reduced ownership due to improved road conditions from fewer PCU. However, trip time increases as more people shift to public transit, involving waiting time and walking time. Interestingly, the monetary cost initially rises and then falls with reduced ownership. For instance, at a 50% ownership reduction (ban=0.5), the cost for all ODs increases from the base 13.21 to 13.29 yuan/trip. However, with a further reduction to a 70% reduction, it drops slightly to 13.28 yuan/trip. This fluctuation is due to changes in trip time and mode expenses. Initially, it rises due to longer trip time, but decrease as solo driving, typically more expensive, declines.

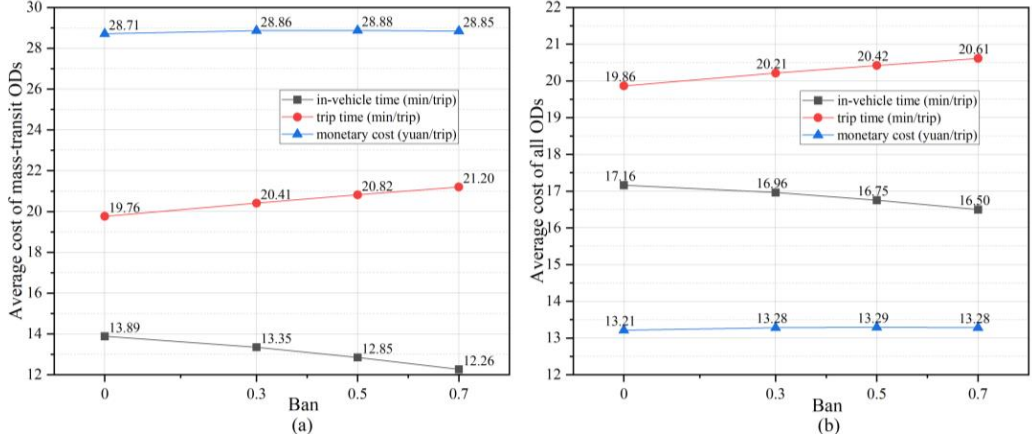

**Figure 6    Time costs and monetary costs under ownership bans**

Four pollutants are observed: $NO_X$, $PM_{2.5}$, CO, and $CO_2e$. Figure 7 presents the distribution of the sum of the four emissions per road link (SPRL) under various disturbances after system stabilization, with a reference line for median SPRL under the base condition. The accompanying table details each kind of emissions in the road network.

This section focuses on the impact of ownership changes on emissions. Reduced ownership leads to lower emissions across the network and SPRL. Specifically, under the base condition, average SPRL is 459747.6g, with CO:1785.0g, $NO_X$: 66.0g, $PM_{2.5}$: 11.6g, $CO_2e$: 457885.0g, the highest SPRL is 1541424.8g, the lowest is 1421.8g, and the median is 353781.3g. A 30% ownership reduction lowers the average SPRL to is 435986.1g, a 5.2% reduction. And the maximum, minimum, and median SPRL are 1449010.4g, 947.8g, 345132.2g, respectively. With a 70% reduction, average SPRL drops to 393,071.4g, a 14.5% reduction, with maximum link emissions down by 15.1%, minimum by 66.7%, and the median by 10.7%.

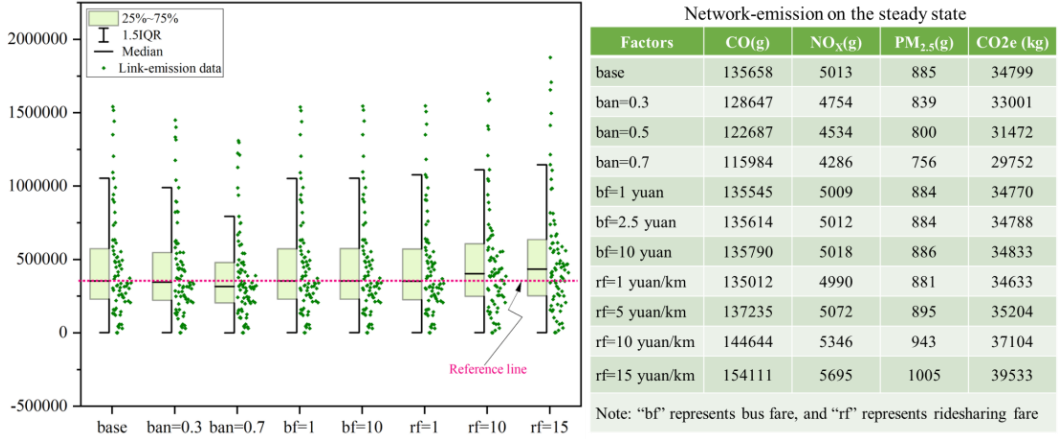

Figure 7　Sensitivity analysis of emissions

### 4.2. Bus fare

The base bus fare is 5 yuan. This section examines the effects of adjusting the bus fare to 1, 2.5, and 10 yuan while keeping other parameters constant (i.e., metro fare at 5 yuan and ridesharing fare at 2 yuan/km). Table 4 shows that as the bus fare increases from 1 to 10 yuan, bus trips decline, while metro and ridesharing trips rise. Figure 8 shows similar modal split trends for both mass-transit and all ODs, with more travelers opting for ridesharing and solo driving, leading to an increase in PCU.

Additionally, it highlights bus and metro competition. Table 4 reveals that when the bus fare is lower than the metro fare, buses are preferred. However, when the bus fare exceeds the metro fare, metro trips rise. A bus fare of 10 yuan leads to a 22-trip (0.6%) increase in metro use, with smaller rises in solo driving (10 trips) and ridesharing (14 trips), showing a stronger shift toward metro use compared to other modes.

Table 4　Modal split and PCU for various bus fare values

| Bus fare (yuan) | All ODs | | | | | Mass-transit ODs | | | | |
|---|---|---|---|---|---|---|---|---|---|---|
| | solo | bus | metro | Ridesharing (driver & pax) | PCU | solo | bus | metro | Ridesharing (driver & pax) | PCU |
| 1 | 11346 | 3390 | 3344 | 7988 | 15430 | 2258 | 3390 | 3344 | 3004 | 3850 |
| 2.5 | 11352 | 3377 | 3350 | 7994 | 15439 | 2264 | 3377 | 3350 | 3010 | 3859 |
| 5 | 11359 | 3354 | 3360 | 7998 | 15448 | 2271 | 3354 | 3360 | 3014 | 3868 |
| 10 | 11369 | 3310 | 3382 | 8012 | 15465 | 2281 | 3310 | 3382 | 3028 | 3885 |

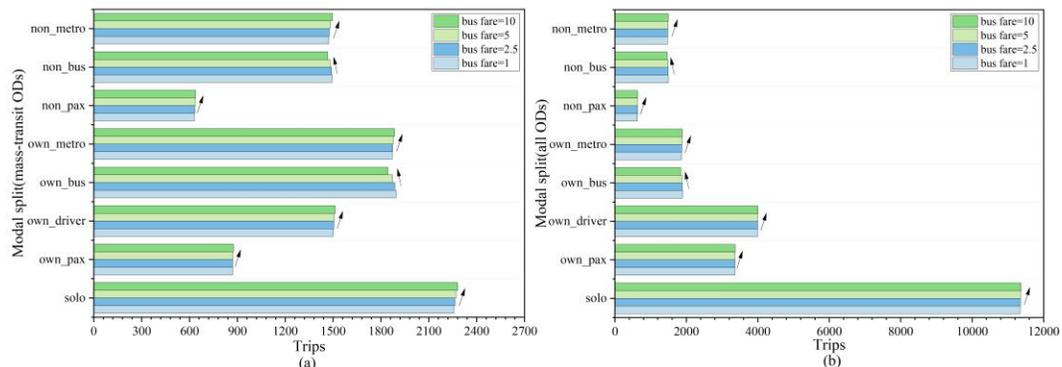

Figure 8　Moda split of different groups under different bus fares

Figure 9 shows that in-vehicle time, trip time, and monetary cost rise with bus fares. However, the rate of increase gradually slows down. For instance, from 1 to 5 yuan, the monetary cost increases by 1.03% per yuan, while from 5 to 10 yuan, it rises by only 0.96%. Notably, the monetary cost sees the largest increase among the three costs, escalating by 9.1% from 1 to 10 yuan, while in-vehicle and trip time increases are marginal. This suggests that bus fare adjustments have minimal impact on traffic efficiency. Furthermore, as depicted in Figure 7, the median and mean values of roadway emissions increase slightly, by about 0.1%, as bus fares rise from 1 to 10 yuan.

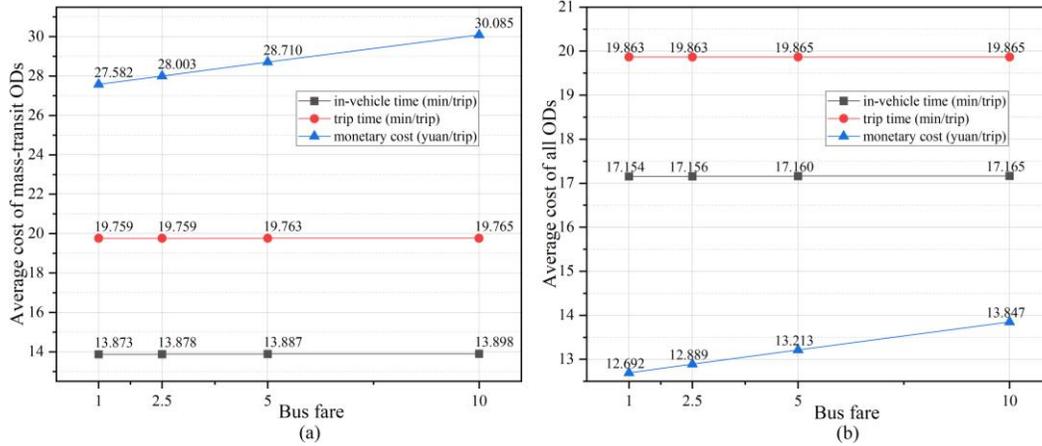

Figure 9　Time costs and monetary costs under different bus fares

### 4.3. Ridesharing fare

The base ridesharing fare is 2 yuan/km, and we analyzed adjustments to 1, 10, and 15 yuan/km. Figure 10 illustrates how fares from 1 to 15 yuan/km impact solo driving, ridesharing, mass transit, and PCU. As fares rise, mass transit demand decreases. Solo driving and ridesharing follow opposite trends, visible in both mass-transit ODs and all ODs. In mass-transit ODs (Figure 10 (a)), solo driving initially drops, then rises after 10 yuan/km, while ridesharing increases and then declines. In all ODs (Figure 10 (b)), solo driving increases and ridesharing decreases without reversal. This results in differing PCU trends. For mass-transit ODs, PCU initially drops from 3893 to 3810 but rebounds to 4144 as fares exceed 5 yuan/km. For all ODs, PCU consistently increases.

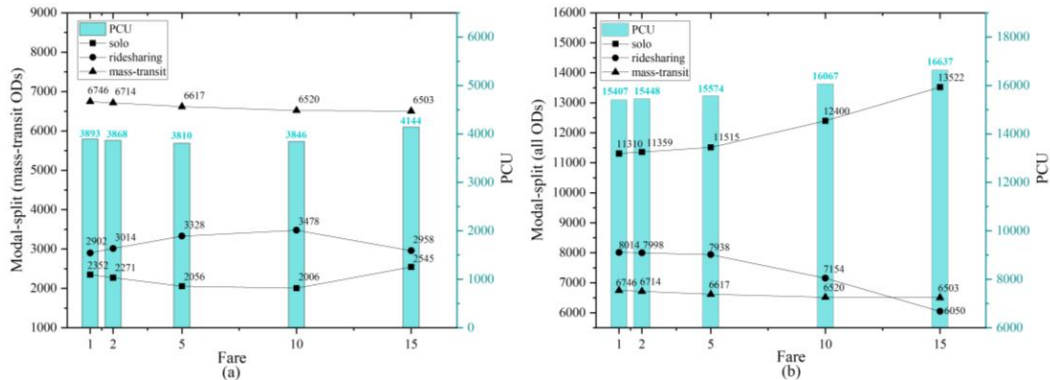

Figure 10　Solo driving, ridesharing, mass-transit, and PCU under various ridesharing fares

Figure 11 illustrates differences in modal choice between car owners and non-owners for both OD sets. In mass-transit ODs (Figure 11 (a)), ridesharing trips (both pax and drivers) first rise with increasing fares, then decline. Public transit trips among non-owners and solo driving trips first drop and then rise. When fares rise from 1 to 10 yuan/km, non-owners favor ridesharing over mass transit due to cost-effectiveness, while more drivers join ridesharing for profit, reducing solo driving and increasing the ridesharing availability. At 15 yuan/km, ridesharing becomes too costly, pushing non-owners back to public transit.

Figure 11 (b) shows that, for all ODs, as fares increase, car owners' use of ridesharing and public transit declines, while solo driving rises. Non-owners' demand for ridesharing and public transit also fluctuates similarly to mass-transit ODs.

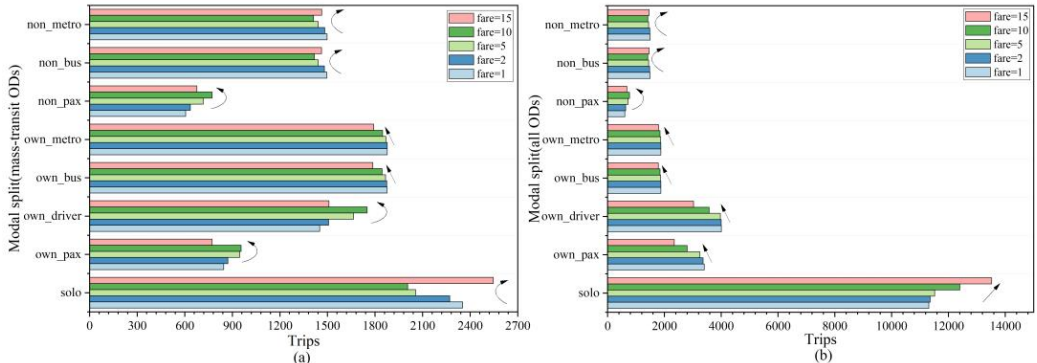

**Figure 11    Moda split of different groups under the ridesharing fares**

As shown in Figure 12, the monetary costs for mass transit ODs and all ODs with increasing the fares, but at different rates across different fare ranges in both OD sets. For instance, in mass-transit ODs, the cost increases fastest between 5 and 10 yuan/km, at 0.4 yuan/trip for each 1 yuan/km fare increase. In contrast, the cost rises by only 0.04 and 0.02 yuan/trip in the ranges of 1-5 and 10-15 yuan/km, respectively. Meanwhile, time costs differ between the two OD sets: for mass-transit ODs, both in-vehicle and trip time initially decline, then rise as fares increase from 1 to 15 yuan/km (Figure 12 (a)), For all ODs, time costs consistently rise (Figure 12 (b)). In general, excessive fares result in higher time costs for travelers.

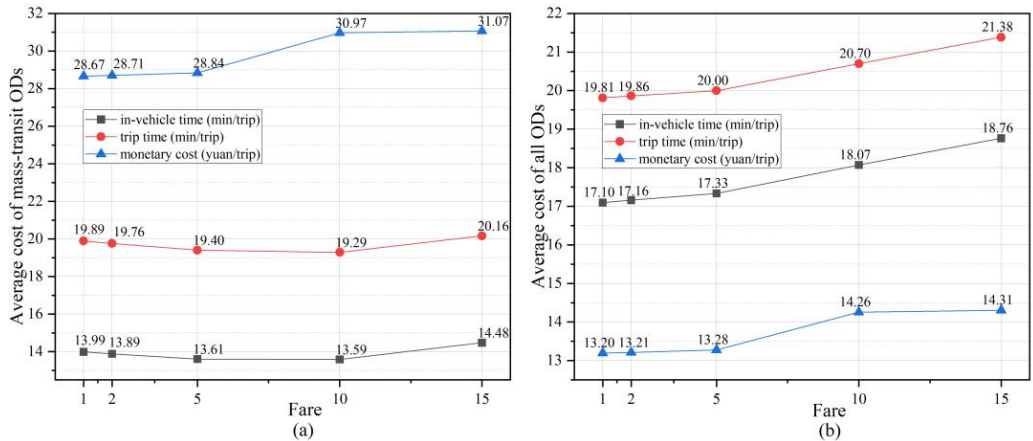

**Figure 12    Time costs and monetary costs under the ridesharing fares**

Figure 7 shows that as ridesharing fares increase, emissions of CO, $NO_X$, $PM_{2.5}$, and $CO_2e$ rise. When ridesharing fare is 1 yuan/km, the average SPRL is 457557.3g, including CO: 1776.5g, $CO_2e$: 455.7kg, $NO_X$: 65.7g, $PM_{2.5}$: 11.6g, with a maximum of 1546164.0g, minimum of 1421.8g, and median of 352359.5g. At 15 yuan/km, average emissions increase to 522,283.2g, a 14.1% rise. The maximum emission rises by 21.4%, minimum by 33.3%, and median by 23.4%.

## 5.  Policy implication

Based on the sensitivity analysis involving transportation demand management policies, specifically vehicle restrictions and pricing strategies, we observe that they have varying impacts on modal choices, traffic conditions, and traffic emissions in MT system. These findings provide important insights for policy formulation and evaluation:

(1) The degree of impact differs depending on the specific measure employed. In the case studied, changes in bus fares have the least impact on vehicle emissions and travel time compared to vehicle restrictions and ridesharing fare adjustments.

(2) Each measure has its trade-offs. For instance, reducing the bus fare to 1 yuan results in the lowest monetary cost in the MT system, but does not significantly reduce emissions. Meanwhile, restricting vehicle proves to be more effective in reducing emissions, despite it increases the average monetary cost and travel time.

(3) The effects of diverse policies show various trends. For instance, as restricting more cars or ridesharing fare increase, there is an inflection point in cost changes, while costs under other measures show monotonic changes. Additionally, the rate of increase in the average monetary cost varies across different ranges of the rising ridesharing fares.

(4) The same policy can have different effects on various aspects of the system. For instance, as ridesharing fares increase, inflection points in PCU changes and modal split (on mass-transit ODs) occur at different fare levels—5 yuan/km and 10 yuan/km, respectively. Excessive ridesharing fares not only increase time costs for travelers but also lead to higher emissions.

Therefore, when implementing the MT system with relevant control policies, it's necessary to balance the advantages and disadvantages of each measure according to urban or regional development needs.

Finally, it is crucial to consider the differential impacts of these policies on urban or regional development and social equity. These impacts vary across populations (vehicle owners vs. non-owners) or across regions (mass transit availability). For instance, vehicle restriction and ridesharing fare adjustments affect mode choice differently between owners and non-owners. Furthermore, the impact of ridesharing fare on various OD sets, such as mass-transit OD and all OD, also varies. For a more detailed insight, please refer to Section 4. The disparity is also evident in road links. Figure 13 depicts the distribution of SPRL on the network at steady state on base condition. Roads used by public transit lines, marked by wide orange lines, experience higher pollution due to increased traffic, even assuming zero emissions from transit vehicles, posing health risks to travelers and residents along these routes. Emphasizing social equity is recommended when formulating relevant policies and urban planning initiatives.

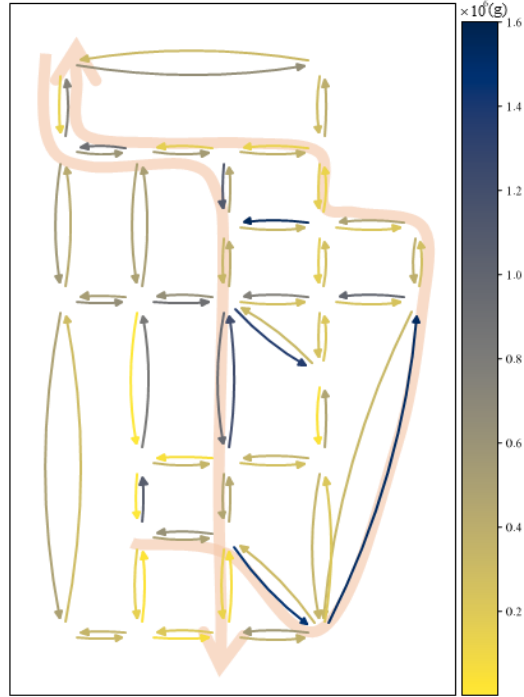

**Figure 13    Distribution of SPRL on Sioux-Falls network at steady state without disturbance**

## 6. Conclusion and outlook

This paper presents the development of a multimodal traffic model aimed at depicting a MT system with ridesharing and offers suggestions for the implementation of various policies or measures. The proposed model integrates the path-based day-to-day traffic dynamics with an optimization ridesharing matching method, including two groups of travelers (vehicle owners and non-owners) and five types of travel modes (i.e., solo driving, ridesharing driving, ridesharing taking, bus taking, and metro taking). Interaction between buses and cars on the road is considered. Results reveal that vehicle restrictions and pricing strategies have varying impacts on modal choices, traffic conditions, and vehicle emissions in MT systems. Meanwhile, the impact of these policies varies across populations and regions and may lead to social equity issues. It implies the necessity of making trade-offs based on the benefits and shortcomings of each measure and emphasizes the importance of balancing equity to improve transportation quality and ensure environmental sustainability.

There are limitations in this study, however, several improvements can be addressed in future work: in this paper, we do not test large-scale networks due to computing power limitations which we plan to address in future studies. More complex travel patterns and behaviors should be considered to better reflect real-world demands. Currently, this paper's ridesharing matching model only accommodates 1-to-1 service between drivers and pax, without addressing the 1-to-many scenario. However, the optimization model presented can be extended to incorporate the 1-to-many case by modifying the constraints, offering a promising direction for future research. Additionally, further exploration is needed in the behaviors of cancellations and mode adjustments during ridesharing matching in multimodal traffic model. Moreover, this paper simplifies the modeling of public transportation, omitting considerations such as bus capacity constraints and transfer connection. Future research could enhance the public transit component of the multimodal traffic model by incorporating elements such as bus departure frequencies, carrying capacities, passenger arrival distributions, and public transport interchanges. Similarly, the modeling of emissions could be improved by accounting for various traffic operations. Finally, only a subset of the model's parameters is involved in sensitive analysis, providing policy suggestions on pricing and driving restriction. In the future, other parameters (e.g., VOT) could be analyzed or congestion toll could be added to cost calculations to offer more comprehensive policy guidance.